\documentclass[prd,aps,eqsecnum,amsmath,floatfix,nofootinbib,preprint,tightenlines]{revtex4}

\usepackage{latexsym}
\usepackage{graphicx}
\usepackage{multirow}
\usepackage[dvipsnames]{xcolor}

\def\floatcaption#1#2{ \caption{#2 \label{#1}} }

\def\bibi{\bibitem}

\def\a{\alpha}
\def\b{\beta}

\def\d{\delta}
\def\g{\gamma}

\def\m{\mu}
\def\n{\nu}
\def\o{\omega}
\def\p{\pi}                     
\def\r{\rho}                    
\def\t{\tau}

\def\z{\zeta}

\def\G{\Gamma}

\def\L{\Lambda}

\def\P{\Pi}

\def\cbo{{\,\raise-.15ex\Sc [\,}}                       

\def\ddt#1{{\buildrel {\hbox{\LARGE .\kern-2pt.}} \over {#1}}}

\def\ie{\mbox{\it i.e.}}
\def\eg{\mbox{\it e.g.}}

\def\half{{1\over 2}}
\def\Re{{\rm Re\,}}

\long \def \blockcomment #1\endcomment{}

\def\eg{{\it e.g.}}

\def\seef{{\it cf.}}

\def\ta{\tilde{a}}
\def\hB{\hat{B}}

\blockcomment
.....
\endcomment

\begin{document}

\begin{center}
\vspace{10mm}
\begin{boldmath}
{\large\bf 
Chiral extrapolation of the leading hadronic 
contribution to the muon anomalous magnetic moment
}
\end{boldmath}

\vspace{3ex}
Maarten~Golterman,$^a$
 Kim~Maltman,$^{b,c}$ Santiago~Peris$^d$%
\\[0.1cm]
{\it
\null$^a$Department of Physics and Astronomy\\
San Francisco State University, San Francisco, CA 94132, USA\\
\null$^b$Department of Mathematics and Statistics\\
York University,  Toronto, ON Canada M3J~1P3\\
\null$^c$CSSM, University of Adelaide, Adelaide, SA~5005 Australia\\
\null$^d$Department of Physics and IFAE-BIST, Universitat Aut\`onoma de Barcelona\\
E-08193 Bellaterra, Barcelona, Spain}
\\[6mm]
{ABSTRACT}
\\[2mm]
\end{center}
\begin{quotation}
A lattice computation of the leading-order hadronic contribution to 
the muon anomalous magnetic moment can potentially help reduce the 
error on the Standard Model prediction for this quantity, if sufficient 
control of all systematic errors affecting such a computation
can be achieved. One of these systematic errors is that associated
with the extrapolation to the physical pion mass from values on the 
lattice larger than the physical pion mass. We investigate this 
extrapolation assuming lattice pion masses in the range of 200 to 400~MeV 
with the help of two-loop chiral perturbation theory, and find that such 
an extrapolation is unlikely to lead to control of this systematic error 
at the 1\% level. This remains true even if various tricks to improve the 
reliability of the chiral extrapolation employed in the literature 
are taken into account. In addition, while chiral perturbation theory also
predicts the dependence on the pion mass of the leading-order 
hadronic contribution to the muon anomalous magnetic moment as the chiral 
limit is approached, this prediction turns out to be of no practical use, 
because the physical pion mass is larger than the muon mass that sets 
the scale for the onset of this behavior.
\end{quotation}

\vfill
\eject
\setcounter{footnote}{0}

\newpage
\section{\label{introduction} Introduction}
Recently, there has been an increasing interest in a high-precision lattice 
computation of the leading-order hadronic vacuum polarization (HVP) 
contribution to the muon anomalous magnetic moment, $a_\mu^{\rm HLO}$. We 
refer to Ref.~\cite{Wittiglat16} for a recent review, and to 
Refs.~\cite{AB2007,Fengetal,UKQCD,DJJW2012,FHHJPR,HPQCDstrange,BE,HPQCDdisconn,RBCdisconn,HPQCD,RBCstrange,BMW16} for efforts in this direction.
The aim is to use the methods of lattice QCD to arrive at a value for 
$a_\mu^{\rm HLO}$ with a total error of one-half to one percent or 
less. Such a result would help solidify, and eventually reduce, 
the total error on the Standard-Model value of the total muon anomalous 
magnetic moment $a_\m$, which is currently dominated by the error on the 
HVP contribution. This desired accuracy requires both a high-statistics 
computation of the HVP, in particular at low momenta (or, 
equivalently, at large distance), as well as a theoretically clean
understanding of the behavior of the HVP as a function of the Euclidean 
squared-momentum $Q^2$ \cite{Pade,taumodel,strategy}, in order 
to help in reducing systematic errors. In addition, it is important to 
gain a thorough understanding of various other systematic errors 
afflicting the computation, such as those caused by a 
finite volume \cite{BM,Mainz,ABGPFV,Lehnerlat16}, scale setting uncertainties,
and the use of  lattice ensembles with light quark masses larger than 
their physical values.  Isospin breaking, electromagnetic effects, 
the presence of dynamical charm and the contribution of quark-disconnected 
diagrams also all enter at the percent level, and thus also have to 
be understood quantitatively with sufficient precision.

In this article, we consider the extrapolation of $a_\mu^{\rm HLO}$ from
heavier than physical pion masses to the physical point, with the help 
of chiral perturbation theory (ChPT). While lattice computations are now 
being carried out on ensembles with light quark masses chosen such that 
the pion mass is approximately physical, a number of computations obtain 
the physical result via extrapolation from heavier pion masses, while 
others incorporate results from heavier pion masses 
in the fits used to convert near-physical-point to actual-physical-point 
results. The use of such heavier-mass ensembles has a potential advantage
since increasing pion mass typically corresponds to decreasing statistical 
errors on the corresponding lattice data.
It is thus important to investigate the reliability of extrapolations of
$a_\mu^{\rm HLO}$ from such heavier masses, say,
$m_\pi\approx 200$~MeV or above, to the physical pion mass.

The leading hadronic contribution is given in terms of the hadronic 
vacuum polarization, and can be written as \cite{ER,TB2003}
\begin{subequations}
\label{amu}
\begin{eqnarray}
a_\m^{\rm HLO}&=&-4\a^2\int_0^\infty \frac{dQ^2}{Q^2}\,w(Q^2)\,
\P_{\rm sub}(Q^2)\ ,\label{amua}\\
w(Q^2)&=&\frac{m_\m^2Q^4Z^3(Q^2)(1-Q^2Z(Q^2))}{1+m_\m^2Q^2Z^2(Q^2)}\ ,
\label{amub}\\
Z(Q^2)&=&\frac{\sqrt{Q^4+4m_\m^2Q^2}-Q^2}{2m_\m^2Q^2}\ ,\label{amuc}\\
\P_{\rm sub}(Q^2)&=&\P(Q^2)-\P(0)\ ,\label{amud}
\end{eqnarray}
\end{subequations}
where $\P(Q^2)$, defined by
\begin{equation}
\label{vacpol}
\P_{\m\n}(Q)=(Q^2\d_{\m\n}-Q_\m Q_\n)\P(Q^2)\ ,
\end{equation}
is the vacuum polarization of the electromagnetic (EM) current,
$\a$ is the fine-structure constant, and $m_\m$ is the 
muon mass.

If we wish to use ChPT, we are restricted to considering only the 
low-$Q^2$ part of this integral, because the ChPT representation of 
$\P_{\rm sub}(Q^2)$ is only valid at sufficiently low values of $Q^2$ 
(as will be discussed  in more detail in Sec.~\ref{comparison}). In view of 
this fact, we will define a truncated $a_\m^{\rm HLO}(Q^2_{max})$:
\begin{equation}
\label{amutrunc}
a_\m^{\rm HLO}(Q^2_{max})=-4\a^2\int_0^{Q^2_{max}} \frac{dQ^2}{Q^2}\,w(Q^2)\,
\P_{\rm sub}(Q^2)\ ,
\end{equation}
and work with $Q^2_{max}$ small enough to allow for the use of ChPT.

In order to check over which $Q^2$ range we can use ChPT, we need data to 
compare with. Here, we will compare to the subtracted vacuum polarization 
obtained using the non-strange $I=1$ hadronic vector spectral function 
measured in $\t$ decays by the ALEPH collaboration \cite{ALEPH13}. Of 
course, in addition to the $I=1$ part $\P^{33}(Q^2)$, the 
vacuum polarization also contains an
$I=0$ component $\P^{88}(Q^2)$ (and, away from the isospin limit,
a mixed isovector-isoscalar component as well). In the isospin limit,\footnote{In Ref.~\cite{ABT}, $\P^{33}$ and $\P^{88}$ are denoted as
$\P^{(1)}_{V\p}$ and $\P^{(1)}_{V\eta}$, respectively.}
\begin{equation}
\label{EMVP}
\P_{\rm EM}(Q^2)=\half\,\P^{{33}}(Q^2)+\frac{1}{6}\,\P^{{88}}(Q^2)\ ,
\end{equation}
where $\P^{33}$ and $\P^{88}$ are defined from the octet vector currents
\begin{eqnarray}
\label{currents}
V_\m^{3}&=&\frac{1}{\sqrt{2}}\left(V_\m^{uu}-V_\m^{dd}\right)\ ,\\
V_\m^{8}&=&\frac{1}{\sqrt{6}}\left(V_\m^{uu}+V_\m^{dd}-2V_\m^{ss}\right)\ ,\nonumber
\end{eqnarray}
with the EM current given by
\begin{equation}
\label{emcurrent}
V^{\rm EM}_\m=\frac{1}{\sqrt{2}}\left(V^{3}_\m+
\frac{1}{\sqrt{3}}\,V^{8}_\m\right)\ .
\end{equation}
Here $V_\m^{uu}=\overline{u}\g_\m u$, $V_\m^{dd}=\overline{d}\g_\m d$ and $V_\m^{ss}=\overline{s}\g_\m s$.
The quantity we will thus primarily consider in this article is\footnote{This 
quantity, at varying values of $Q^2_{max}$, was considered before in
Refs.~\cite{taumodel,strategy}.}
\begin{equation}
\label{amutilde}
\ta_\m(Q^2_{max})=-4\a^2\int_0^{Q^2_{max}} \frac{dQ^2}{Q^2}\,w(Q^2)\,
\P^{{33}}_{\rm sub}(Q^2)\ ,
\end{equation}
where we will choose $Q^2_{max}=0.1$~GeV$^2$ (\seef\ Sec.~\ref{comparison} below).
In Sec.~\ref{EM} we will consider also the inclusion of the $I=0$ contribution.

It is worth elaborating on why we believe the quantity 
$\ta_\m(Q^2_{max}=0.1~\mbox{GeV}^2)$ will be useful for studying the 
extrapolation to the physical pion mass, in spite of the fact that it
constitutes only part of $a_\m^{\rm HLO}$. First, the $I=1$ threshold is 
$s=4m_\p^2$, while that for $I=0$ is $s=9m_\p^2$.\footnote{In fact, 
to NNLO in ChPT, the threshold is $s=4m_K^2$.} This suggests that the $I=1$ 
part of $a_\m^{\rm HLO}$ should dominate the chiral behavior. 
Second, from the dispersive representation, it is clear that the 
relative contributions to $\P_{\rm sub}(Q^2)$ from the region near the 
two-pion threshold are larger at low $Q^2$ than they are at high $Q^2$. 
Contributions to $a_\m^{\rm HLO}$ from the low-$Q^2$ part of the integral 
in Eq.~(\ref{amua}) are thus expected to be relatively more sensitive to 
variations in the pion mass than are those from the rest of the integral. 
The part of the integral below $Q^2_{max}=0.1$~GeV$^2$, moreover, yields 
about 80\% of $a_\m^{\rm HLO}$. We thus expect a study of the chiral 
behavior of $\ta_\m(Q^2_{max}=0.1~\mbox{GeV}^2)$ to provide important 
insights into the extrapolation to the physical pion mass.  This
leaves out the contribution from the integral above $0.1$~GeV$^2$, 
which can be accurately computed directly from the lattice 
data using a simple trapezoidal rule evaluation \cite{taumodel,strategy}.
Its pion mass dependence is thus not only expected to be milder, for the
reasons given above, but also to be amenable to a direct study using 
lattice data. In light of these comments, it seems to us highly unlikely 
that adding the significantly smaller ($\sim 20\%$) $Q^2>0.1$~GeV$^2$ 
contributions, with their weaker pion mass dependence, could produce
a complete integral with a significantly reduced sensitivity to the 
pion mass.

This paper is organized as follows. In Sec.~\ref{chpt} we collect the needed
expressions for the HVP to next-to-next-to-leading order (NNLO) in 
ChPT, and derive a formula for the dependence of $a_\m^{\rm HLO}$ on 
the pion mass in the limit $m_\pi\to 0$. In Sec.~\ref{ALEPH} we compare the
$I=1$ ChPT expression with the physical $\P^{{33}}_{\rm sub}(Q^2)$, 
constructed from the ALEPH data, and argue that 
$\ta_\m(Q^2_{max}=0.1~\mbox{GeV}^2)$ can be reproduced to an accuracy of 
about 1\% in ChPT. Section~\ref{extrapolation} contains the study of the 
extrapolation of $\ta_\m(Q^2_{max}=0.1~\mbox{GeV}^2)$ computed at pion masses 
typical for the lattice, also considering various tricks that have been
considered in the literature to modify $a_\m^{\rm HLO}$ at larger
pion mass in such a way as to weaken the pion-mass-dependence of the result 
and thus improve the reliability of the thus-modified chiral extrapolation.
We end this section with a discussion of the inclusion of the $I=0$ part.
We present our conclusions in Sec.~\ref{conclusion},
and relegate some technical details to an appendix.

\section{\label{chpt} The vacuum polarization in chiral perturbation theory}
In this section we collect the NNLO expressions for $\P^{{33}}(Q^2)$ and 
$\P^{{88}}(Q^2)$ as a function of Euclidean $Q^2$, summarizing the results 
of Refs.~\cite{GK95,ABT}. Using the conventions of Ref.~\cite{ABT}, one has
\begin{eqnarray}
\label{Pi1}
\P^{{33}}_{\rm sub}(Q^2)&=&-8\hB(Q^2,m_\p^2)-4\hB(Q^2,m_K^2)\\
&&+\frac{16}{f_\p^2}\,L_9^r \,Q^2\left(2B(Q^2,m_\p^2)+B(Q^2,m_K^2)
\right)\nonumber\\
&&-\frac{4}{f_\p^2}\,Q^2\left(2B(Q^2,m_\p^2)+B(Q^2,m_K^2)
\right)^2\nonumber\\
&&+8C_{93}^rQ^2+C^r(Q^2)^2\ ,
\nonumber
\end{eqnarray}
and
\begin{eqnarray}
\label{Pi0}
\P^{{88}}_{\rm sub}(Q^2)&=&-12\hB(Q^2,m_K^2)\\
&&+\frac{48}{f_\p^2}\,L_9^r\,Q^2B(Q^2,m_K^2)
-\frac{36}{f_\p^2}\,Q^2\left(B(Q^2,m_K^2)\right)^2
\nonumber\\
&&+8C_{93}^rQ^2+C^r(Q^2)^2\ ,
\nonumber
\end{eqnarray}
where $B(Q^2,m^2)=B(0,m^2)+\hB(Q^2,m^2)$ is the subtracted
standard equal-mass,
two-propagator, one-loop integral, with
\begin{eqnarray}
\label{B}
B(0,m^2)&=&\frac{1}{192\p^2}\left(1+\log{\frac{m^2}{\m^2}}\right)\ ,\\
\hB(Q^2,m^2)&=&\frac{1}{96\p^2}
\left(\left(\frac{4 m^2}{Q^2}+1\right)^{3/2}  \mbox{coth}^{-1}
   \sqrt{1+\frac{4m^2}{Q^2}}- 
   \frac{4m^2}{Q^2}-{\frac{4}{3}}\right)\ ,\nonumber
\end{eqnarray}
and the low-energy constants (LECs) $L_9^r$ and $C_{93}^r$ are renormalized
at the scale $\m$, in the ``$\overline{MS}+1$'' scheme 
employed in Ref.~\cite{ABT}. Note that these are 
the only two LECs appearing in the subtracted versions of
the $I=1$ and $I=0$ non-strange vacuum polarizations to NNLO.

As in Ref.~\cite{strategy}, we have added an analytic NNNLO term, 
$C^r(Q^2)^2$, to $\P^{33}_{\rm sub}(Q^2)$ and $\P^{88}_{\rm sub}(Q^2)$
in order to improve, in the $I=1$ case, the agreement with 
$\P^{{33}}_{\rm sub}(Q^2)$
constructed from the ALEPH data, as we will see in Sec.~\ref{ALEPH} below.   
Such a contribution, which would first appear at NNNLO in the chiral
expansion, and be produced by six-derivative terms in the NNNLO Lagrangian,
will necessarily appear with the same coefficient in both the $I=0$ and $I=1$ 
polarizations at NNNLO.\footnote{Singling out the $C^r(Q^2)^2$ term 
from amongst the full set of NNNLO contributions introduces a phenomenological
element to our extended parametrization.
As noted in Ref.~\cite{strategy}, the
fact that one finds $C_{93}^r$ to be dominated by the contribution
of the $\rho$ resonance leads naturally to the expectation that the
next term in the expansion of the $\rho$ contribution at low $Q^2$,
which has precisely the form $C^r (Q^2)^2$, should begin to become 
numerically important already for $Q^2$ as low as $\sim 0.1\ {\rm GeV}^2$.}
Following Ref.~\cite{strategy}, we will refer to the expressions with $C^r=0$ 
as ``NNLO,'' and with the $C^r$ term included as ``NN$^\prime$LO.''   

From these expressions, it is clear that the chiral behavior of 
$a_\m^{\rm HLO}$, which we expect to be primarily governed
by the pion, rather than kaon, contributions, will be dominated by 
the $I=1$ component. In fact, in the limit that $m_\p\to 0$ with 
$m_\m$ fixed, we find
\begin{eqnarray}
\label{amufinal}
a_\m^{I=1}&\equiv&\ta_\m(Q^2_{max}=\infty)=\\
&&\frac{\a^2}{12\p^2}\left(-\log{\frac{m_\p^2}{m_\m^2}}
-\frac{31}{6}+3\p^2\sqrt{\frac{m_\p^2}{m_\m^2}}
+O\left(\frac{m_\p^2}{m_\m^2}\log^2{\frac{m_\p^2}{m_\m^2}}\right)\right)\ .
\nonumber
\end{eqnarray}
A derivation of this result is given in App.~\ref{appendix}. 
We note that the scale for the chiral extrapolation is set by the 
muon mass, and thus Eq.~(\ref{amufinal}) applies to the region $m_\p\ll m_\m$.
Therefore, this result is unlikely to be of much practical value. 
Indeed, our tests in Sec.~\ref{extrapolation} below will
confirm this expectation. 
 
\begin{boldmath}
\section{\label{ALEPH} Comparison with ALEPH data for hadronic $\t$ decays}
\end{boldmath}
In this section, we will construct $\P^{{33}}_{\rm sub}(Q^2)$ from the 
non-strange, vector spectral function $\r^{{33}}_V$ measured by ALEPH in 
hadronic $\t$ decays \cite{ALEPH13}, using the once-subtracted 
dispersion relation
\begin{equation}
\label{dispPiV}
\P^{{33}}_{\rm sub}(Q^2)=-Q^2\int_{4m_\p^2}^\infty ds\,
\frac{\r^{{33}}_V(s)}{s(s+Q^2)}\ .
\end{equation}
Since the spectral function is only measured for $s\le m_\t^2$, this is not
entirely trivial, and we will describe our construction in more detail in 
Sec.~\ref{data}. We then compare the data with ChPT in Sec.~\ref{comparison}.

\vskip 0.8cm
\begin{boldmath}
\subsection{\label{data} $\P^{{33}}_{\rm sub}(Q^2)$ from ALEPH data}
\end{boldmath}
In order to construct $\r^{{33}}_V(s)$ for $s>m_\t^2$, we follow 
the same procedure as in the $V-A$ case considered in Refs.~\cite{L10,VmAALEPH}.
For a given $s_{min}\le m_\t^2$, we switch from the data representation of 
$\r^{{33}}_V(s)$ to a theoretical representation given by the sum of the 
QCD perturbation theory (PT) expression $\r^{{33}}_{V,{\rm PT}}(s)$ and 
a ``duality-violating'' (DV) part $\r^{{33}}_{V,{\rm DV}}(s)$ that 
represents the oscillations around perturbation
theory from resonances, and which we model as
\begin{equation}
\label{DV}
\r^{{33}}_{V,{\rm DV}}(s)=e^{-\d_V-\g_V s}\sin{(\a_V+\b_V s)}\ .
\end{equation}
The perturbative expression is known to order $\a_s^4$ \cite{PT}, where
$\a_s=\a_s(m_\t^2)$ is the strong coupling. Fits to the ALEPH data determining
the parameters $\a_s$, $\a_V$, $\b_V$, $\g_V$ and $\d_V$ have been 
extensively studied in Ref.~\cite{alphas14}, with the goal of a 
high-precision determination of $\a_s$ from hadronic $\t$ decays. 
Here we will use the values obtained from the FOPT $s_{min}=1.55$~GeV$^2$ 
fit of Table~1 of Ref.~\cite{alphas14},
\begin{eqnarray}
\label{param}
\a_s(m_\t^2)&=&0.295(10)     \ ,\\
\a_V&=&-2.43(94)     \ ,\nonumber\\
\b_V&=&4.32(48)\ \mbox{GeV}^{-2}     \ ,\nonumber\\
\g_V&=&0.62(29)\ \mbox{GeV}^{-2}     \ ,\nonumber\\
\d_V&=&3.50(50)     \ .\nonumber
\end{eqnarray}
The match between the data and theory representations of the spectral 
function in the window $s_{min}\le s\le m_\t^2$ is excellent, and there 
is no discernible effect on $\P^{{33}}_{\rm sub}(Q^2)$ for the values 
of $Q^2$ smaller than $0.2$~GeV$^2$ of interest in the comparison to ChPT
below  if we use a different switch point from data
to theory inside this interval, switch to a CIPT instead an FOPT fit, 
or if we use the parameter values of one of the other optimal
fits in Ref.~\cite{alphas14}.\footnote{This stability is not
surprising since (i) for small $Q^2$, the weight in the dispersive
representation~(\ref{dispPiV}) falls of as $1/s^2$ for larger $s$,
and (ii) the DV and perturbative contributions to $\r^{{33}}_V(s)$ 
are small relative to the leading parton model contribution in the higher-$s$
region where the PT+DV representation is used.}
Results for $\P^{{33}}_{\rm sub}(Q^2)$ 
in the region below $Q^2=0.2$~GeV$^2$, at intervals of $0.01$~GeV$^2$, are
shown in Fig.~\ref{PiVdata}. The errors shown are fully correlated, taking
into account, in particular, correlations between the parameters of 
Eq.~(\ref{param}) and the data.

\begin{figure}[t]
\vspace*{4ex}
\begin{center}
\includegraphics*[width=14cm]{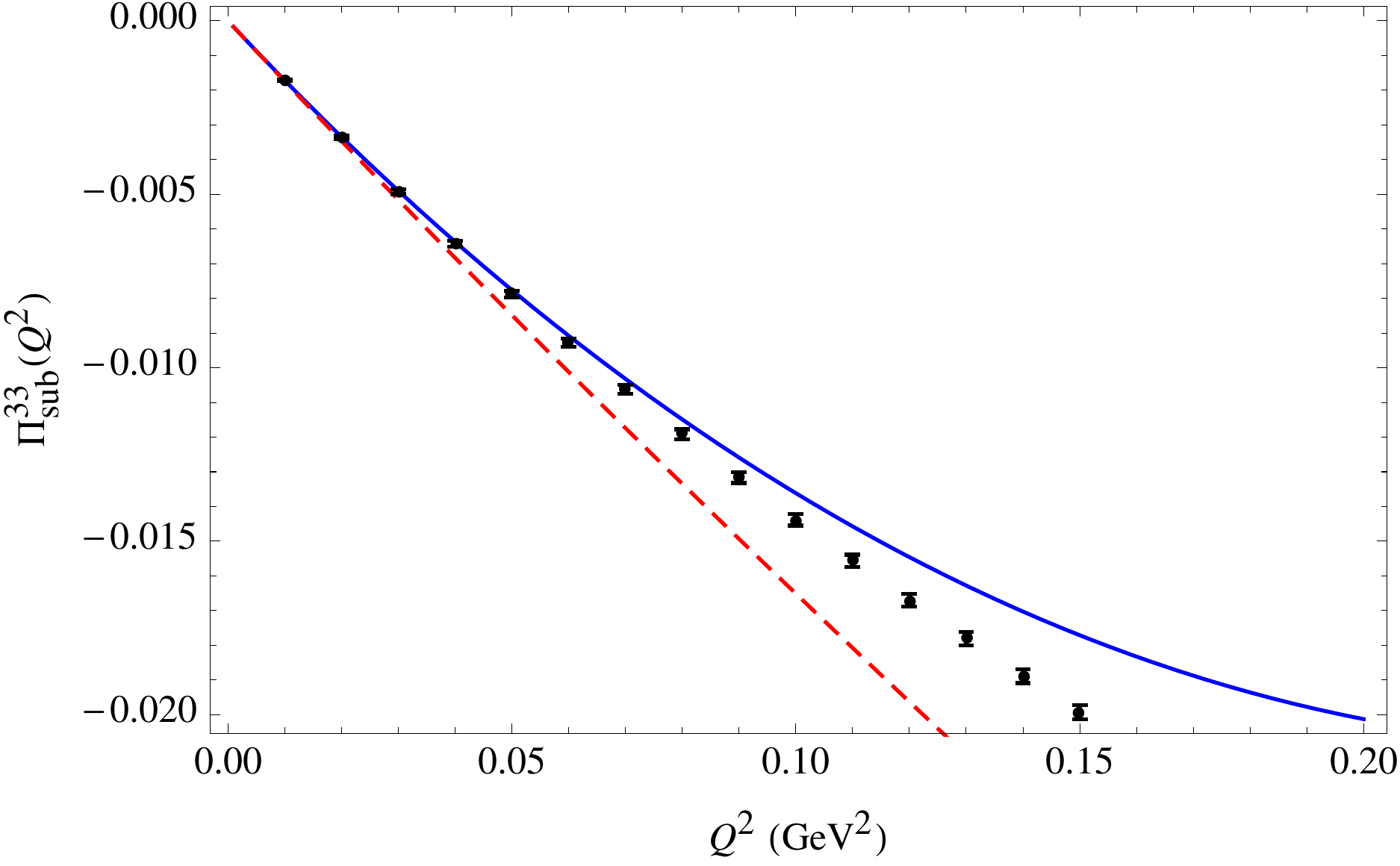}
\end{center}
\begin{quotation}
\floatcaption{PiVdata}%
{\it $\P^{{33}}_{\rm sub}(Q^2)$ as a function of $Q^2$. Black points: data, 
constructed as explained in Sec.~\ref{data}; red (lower) curve: {\rm NNLO} 
ChPT representation with $C^r=0$; blue (upper) curve: {\rm NN$^\prime$LO} 
ChPT representation with $C^r$ as determined from the data. For the 
ChPT representations, see Sec.~\ref{comparison}.}
\end{quotation}
\vspace*{-4ex}
\end{figure}

\vskip 0.8cm
\subsection{\label{comparison} Comparison with ChPT}
In order to compare the data for $\P^{{33}}_{\rm sub}(Q^2)$ with ChPT,
we need values for $L_9^r$, $C^r_{93}$ and $C^r$.  Although, 
in principle, they can all be obtained from a fit to the ALEPH data, in
practice $L_9^r$ and $C^r_{93}$ turn out to be strongly 
anti-correlated, making it difficult to determine these two LECs 
separately from these data. We thus, instead, use an external
value for $L_9^r$ taken from the NNLO analysis of Ref.~\cite{BTL9}:
\begin{equation}
\label{L9}
L^r_9(\m=0.77~\mbox{GeV})=0.00593(43)\ .
\end{equation}
With this value, a fit to the slope and curvature at $Q^2=0$ of
$\P^{{33}}_{\rm sub}(Q^2)$ is
straightforward, and we find
\begin{eqnarray}
\label{Cs}
C^r_{93}(\m=0.77~\mbox{GeV})&=&-0.0154(4)\ \mbox{GeV}^{-2}\ ,\\
C^r(\m=0.77~\mbox{GeV})&=&0.29(3)\ \mbox{GeV}^{-4}\ .\nonumber
\end{eqnarray}
The determination of $C^r_{93}$ is new, and will be discussed in more detail
in a forthcoming publication \cite{C93}. Here, we will use the central values
in a comparison between the data and ChPT, in order to see with what
accuracy $\ta_\m(Q^2_{max})$ can be represented in ChPT, as a function of
$Q^2_{max}$.

The two curves in Fig.~\ref{PiVdata} show ChPT representations of
$\P^{{33}}_{\rm sub}(Q^2)$. The blue solid curve corresponds to
NN$^\prime$LO ChPT, employing the values~(\ref{L9}) and~(\ref{Cs}),
the red dashed curve to NNLO ChPT, obtained by dropping the $C^r(Q^2)^2$
contribution from the fitted NN$^\prime$LO result. It is clear that 
allowing for the analytic NNNLO term in ChPT helps improve the agreement 
with the data, even though this falls short of a full NNNLO comparison.

We may now compare values of $\ta_\m(Q^2_{max})$ computed from the
data and from ChPT, as a function of $Q^2_{max}$.  For $Q^2_{max}=0.1$~GeV$^2$
we find
\begin{equation}
\label{01}
\ta_\m(0.1\ \mbox{GeV}^2)=\left\{
\begin{array}{ll}
9.81\times 10^{-8} & \qquad\mbox{data\ ,} \\
9.73\times 10^{-8} & \qquad\mbox{NN$^\prime$LO ChPT}\ ,\\
10.23\times 10^{-8} & \qquad\mbox{NNLO ChPT}\ .
\end{array}\right.
\end{equation}
We do not show errors, because we are only interested in the ChPT values
for $\ta_\m(Q^2_{max})$ as a model to study the pion mass dependence.
However, Eq.~(\ref{01}) shows that NN$^\prime$LO ChPT reproduces the data
value for $\ta_\m(0.1\ \mbox{GeV}^2)$ to about 1\%, and that the addition 
of the $C^r$ term to Eq.~(\ref{Pi1}) improves this agreement from about 4\%. 
With the value of $\ta_\m=\ta_\m(\infty)=11.95\times 10^{-8}$ computed 
from the data we also see that the $Q^2_{\max}=0.1$~GeV$^2$ value amounts 
to 82\% of the full integral. For $Q^2_{max}=0.2$~GeV$^2$ we find, similarly,
\begin{equation}
\label{02}
\ta_\m(0.2\ \mbox{GeV}^2)=\left\{
\begin{array}{ll}
10.96\times 10^{-8} & \qquad\mbox{data\ ,} \\
10.77\times 10^{-8} & \qquad\mbox{NN$^\prime$LO ChPT}\ ,\\
11.61\times 10^{-8} & \qquad\mbox{NNLO ChPT}\ .
\end{array}\right.
\end{equation}
For $Q^2_{\max}=0.2$~GeV$^2$ the presence of $C^r$ improves the 
agreement between the ChPT and data values from about 6\% to about
2\%, and the truncated integral provides 92\% of the full result.

It is remarkable that ChPT does such a good job for $\ta_\m(Q^2_{max})$
for these values of $Q^2_{max}$, and that such low values of $Q^2_{max}$
already represent such a large fraction of the integral~(\ref{amutilde}) 
for $Q^2_{max}=\infty$.   The reason is that the integrand of Eq.~(\ref{amutilde})
is strongly peaked at $Q^2\approx m_\m^2/4=0.0028$~GeV$^2$.
Below, we will use values of $\ta_\m(Q^2_{max})$
computed with $Q^2_{\max}=0.1$~GeV$^2$
for our study of the pion mass dependence.

\begin{boldmath}
\section{\label{extrapolation} Chiral extrapolation of $\ta_\m(Q^2_{max})$}
\end{boldmath}
The pion mass dependence of $\ta_\m(Q^2_{max})$, as it would be computed
on the lattice, has a number of different sources. Restricting ourselves to
$Q^2_{max}=0.1$~GeV$^2$, we can use ChPT to trace these sources. In 
addition to the explicit dependence on $m_\p$ in Eq.~(\ref{Pi1}), $m_K$ and
$f_\p$ also depend on the pion mass.\footnote{We will assume lattice 
computations with the strange quark fixed at its physical mass, and with 
isospin symmetry, in which only the light quark mass (\ie, the average 
of the up and down quark masses) is varied.}   
Although $C_{93}^r$ and $C^r$ represent LECs of the
effective chiral Lagrangian and hence are mass-independent, the
data includes contributions of all chiral orders. Thus, when we
perform fits using the truncated NNLO and NN$^\prime$LO forms,
the resulting LEC values, in general, will become effective
ones, in principle incorporating mass-dependent contributions
from terms higher order in ChPT than those shown in Eq.~(\ref{Pi1}).  
These will, in general, differ from the true mass-independent LECs 
$C_{93}^r$ and $C^r$ due to residual higher-order mass-dependent effects. 
These same effects would also cause the values obtained from analogous 
fits to lattice data for ensembles with unphysical pion mass to differ 
from the true, mass-independent values. We will denote the general 
mass-dependent effective results 
by $C_{93,{\rm eff}}^{r}$ and $C^{r}_{\rm eff}$, 
and model their mass dependence by assuming the fitted values in 
Eq.~(\ref{Cs}) are dominated by the contributions of the $\r$ 
resonance \cite{strategy,ABT}. With
this assumption,
\begin{eqnarray}
\label{CVMD}
C_{93,{\rm eff}}(\m=0.77~\mbox{GeV})&=&-\frac{f_\r^2}{4m_\r^2}\ ,\\
C^{r}_{\rm eff}(\m=0.77~\mbox{GeV})&=&\frac{2f_\r^2}{m_\r^4}\ ,\nonumber
\end{eqnarray}
with $m_\rho$ and $f_\rho$ in general dependent on the pion mass.
We will suppress the explicit $m_\pi$ dependence of $C_{93,{\rm eff}}^r$
and $C^{r}_{\rm eff}$ except where a danger of confusion exists. For
physical light quark mass, with $f_\r\approx 0.2$ and $m_\r=0.775$~GeV, 
we find $C_{93,{\rm eff}}(m_\pi^2)\approx-0.017$~GeV$^{-2}$, and 
$C^{r}_{\rm eff}(m_\pi^2)\approx 0.22$~GeV$^{-4}$.
These values are in quite reasonable
agreement with Eq.~(\ref{Cs}).

On the lattice, one finds that $m_\r$ is considerably more sensitive to 
the pion mass than is $f_\r$ \cite{Fengetal}. We thus model the pion mass 
dependence of $C^r_{93,{\rm eff}}$ and $C^r_{\rm eff}$ by assuming the 
effective $\m=0.77$~GeV values are given by
 \begin{eqnarray}
\label{Ceff}
C^r_{93,{\rm eff}}(m_{\pi ,latt}^2)&=&
C^r_{93,{\rm eff}}(m_\pi^2)\,\frac{m_\r^2}{m_{\r,{\rm latt}}^2}\ ,\\
C^r_{\rm eff}(m_{\pi ,latt}^2)&=&
C^r_{{\rm eff}}(m_\pi^2)\,\frac{m_\r^4}{m_{\r,{\rm latt}}^4}\ .\nonumber
\end{eqnarray}
where $m_{\r,{\rm latt}}$ is the $\r$ mass computed on the lattice.

This strategy allows us to generate a number of fake lattice data for
$\ta_\m(Q^2_{max})$ using ChPT.  For each $m_\pi$ in the range
of interest, the corresponding $m_\rho$ is needed to compute 
$C_{93,{\rm eff}}^r$ and $C_{\rm eff}^r$ via Eqs.~(\ref{Ceff}). This information
is available, over the range of $m_\pi$ we wish to study, for the HISQ
ensembles of the MILC collaboration \cite{MILC}, and we thus use the
following set of values for $m_\p$, $f_\p$, $m_K$ and $m_\r$, 
corresponding to those ensembles:

\vspace{2ex}
\begin{equation}
\label{MILC}
\begin{array}{|c|c|c|c|}
\hline
m_\p\ (\mbox{MeV}) & f_\p\ (\mbox{MeV}) & m_K\ (\mbox{MeV}) & m_\r\ (\mbox{MeV}) \\
\hline
223 & 98 & 514 & 826 \\
262 & 101 & 523 & 836 \\
313 & 104 & 537 & 859^* \\
382 & 109 & 558 & 894 \\
440 & 114 & 581 & 929\\
\hline
\end{array}
\vspace{2ex}
\end{equation}
The statistical errors on these numbers are always smaller than 1\%, except 
for the $\r$ mass marked with an asterisk. In fact, the (unpublished)
MILC value for this $\r$ mass is $834(30)$~MeV. 
Since we are interested in constructing a model, we corrected
this value by linear interpolation in $m_\p^2$ between the two 
neighboring values, obtaining the value $859$~MeV, which is consistent 
within errors with the MILC value. With this correction, $f_\p$, $m_K$ 
and $m_\r$ are all approximately linear in $m_\p^2$.

\vskip 0.8cm
\subsection{\label{ETMC} The ETMC trick}
Before starting the numerical study of our ChPT-based model, we outline a
trick aimed at modifying $a_\m^{\rm HLO}$ results
at heavier pion masses
in such a way as to weaken the resulting pion mass dependence, and thus 
improve the reliability of the extrapolation to the physical pion mass.
The trick, first introduced in Ref.~\cite{Fengetal}, is best explained using an
example. Consider the following very simple vector-meson dominance (VMD)
model for $\P^{{33}}(Q^2)$ \cite{MHA}:
\begin{equation}
\label{VMD}
\P^{{33}}_{\rm VMD,sub}(Q^2)=-\frac{2f_\r^2 Q^2}{Q^2+m_\r^2}-\frac{1}{4\p^2}\log{\left(1+\frac{Q^2}{8\p^2 f_\r^2 m_\r^2}\right)}\ .
\end{equation}
The logarithm is chosen such that it reproduces the parton-model logarithm
while at the same time generating no $1/Q^2$ term for large $Q^2$.
The (simplest version of the) ETMC trick consists of inserting a 
correction factor $m_{\r,{\rm latt}}^2/m_\r^2$ in front of $Q^2$ in the 
subtracted HVP before carrying out the integral over $Q^2$ in Eq.~(\ref{amu}).
If we assume that $f_\r$ does not depend on $m_\p$ 
but $m_\r$ does, it is easily seen that the resulting ETMC-modified 
version of $\P^{{33}}_{\rm VMD,sub}(Q^2)$, $\P^{{33}}_{\rm VMD,sub}
\left( {\frac{m_{\r,{\rm latt}}^2}{m_\r^2}} Q^2\right)$, is completely 
independent of $m_\pi$. With the VMD form known to provide a reasonable 
first approximation to $\P^{{33}}_{\rm sub}(Q^2)$, the application 
of the ETMC trick to actual lattice results is thus expected to produce 
a modified version of $a_\m^{\rm HLO}$ displaying considerably reduced 
$m_\pi$ dependence. In Ref.~\cite{Fengetal} a further change of variable was
performed to shift the modification factor out of the argument of 
the HVP and into that of the weight function, the result being a
replacement of the argument $Q^2$ in $w(Q^2)$ by 
$\left( m_\r^2/m_{\r,{\rm latt}}^2\right) Q^2$. We do not perform
this last change of variable since, in our study, we cut off the integral at
$Q^2=Q^2_{max}$, \seef\ Eqs.~(\ref{amutrunc}) and~(\ref{amutilde}).

In Ref.~\cite{HPQCD} a variant of this trick was used as follows. First, the 
HVP was modified to remove what was expected to be the strongest 
pion mass dependence by subtracting from the lattice version of 
$\P^{{33}}_{\rm sub}(Q^2)$ the NLO pion loop contribution (effectively, 
from our perspective, the first term of Eq.~(\ref{Pi1})), evaluated 
at the lattice pion mass, $m_{\p,{\rm latt}}$. The ETMC rescaling, 
$Q^2\to (m_{\r,{\rm latt}}^2/m_\r^2)Q^2$, was then applied to the 
resulting differences and the extrapolation to physical pion mass 
performed on these results. Finally, the physical mass version of 
the NLO pion loop contribution (again, effectively the first term 
of Eq.~(\ref{Pi1}), now evaluated at physical $m_\p$) was added back
to arrive at the final result for $a_\m^{\rm HLO}$.\footnote{In 
Ref.~\cite{HPQCD} the ETMC rescaling was actually done at the level of 
the moments used to construct Pad\'e approximants for $\P_{\rm sub}(Q^2)$.
The two procedures are equivalent if the Pad\'e approximants converge.}
This sequence of procedures is equivalent, in our language, to 
employing the modified HVP
\begin{equation}
\label{pioncorr}
\P_{\rm sub,corr}^{33}(Q^2)=\P_{\rm sub}^{33}
\left(\frac{m_{\r,{\rm latt}}^2}{m_\r^2}\,Q^2\right)+8\left(
\hB\left(\frac{m_{\r,{\rm latt}}^2}{m_\r^2}\,Q^2,m_{\p,{\rm latt}}^2\right)
-\hB(Q^2,m_\p^2)\right)\ .
\end{equation}
We will refer to this version of the ETMC trick as the HPQCD trick.
 
\begin{boldmath}
\subsection{\label{I=1} The $I=1$ case}
\end{boldmath}
We have generated three ``data'' sets based on the results for 
$\ta_\m\equiv\ta_\m(Q^2_{max}=0.1$~GeV$^2)$, at the five values of $m_\p$ 
given in Eq.~(\ref{MILC}), using Eq.~(\ref{Pi1}) with the effective LECs~(\ref{Ceff}).
One set consists simply of the five unmodified results for $\ta_\m$, 
the other two of the ETMC- and HPQCD-modified version thereof, all 
obtained using Eq.~(\ref{Pi1}) with the effective LECs~(\ref{Ceff}).
We will refer to these three data sets as unimproved, ETMC-improved, and
HPQCD-improved in what follows. To avoid a proliferation of
notation, and since it should cause no confusion to do so, the ETMC- and
HPQCD-modified versions of $\ta_\m$ will also be denoted by
$\ta_\m$ in what follows. 

We performed three fits on each of these three data sets, using the following
three functional forms for the dependence on $m_\p$:
\begin{subequations}
\label{fits}
\begin{eqnarray}
\ta_\m^{\rm quad}&=&Am_{\p,{\rm latt}}^4+Bm_{\p,{\rm latt}}^2+C\hspace{4cm}(\mbox{quadratic})\ ,\label{fitsa}\\
\ta_\m^{\rm log}&=&A\log{(m_{\p,{\rm latt}}^2/m_\p^2)}+Bm_{\p,{\rm latt}}^2+C
\hspace{2.23cm}(\mbox{log})\ ,\label{fitsb}\\
\ta_\m^{\rm inv}&=&\frac{A}{m_{\p,{\rm latt}}^2}+Bm_{\p,{\rm latt}}^2+C
\hspace{4.19cm}(\mbox{inverse})\ .\label{fitsc}
\end{eqnarray}
\end{subequations}
The ``log'' fit is inspired by Eq.~(\ref{amufinal}).   The ``inverse'' fit is essentially that 
used by HPQCD in Ref.~\cite{HPQCD}.   Explicily, without scaling violations, and assuming 
a physical strange quark mass, the HPQCD fit function takes the form
\begin{equation}
\label{HPQCD}
a_\m^{\rm HLO}\left(1+c_\ell\,\frac{\d m_\ell}{\L}+{\tilde c}_\ell\,\frac{\d m_\ell}{m_\ell}
\right)\ ,
\end{equation}
with $\d m_\ell=m_\ell-m_\ell^{\rm phys}$ and $m_\ell$ the average of the up and
down quark masses on the lattice.\footnote{Ref.~\cite{HPQCD} assumed exact isospin
symmetry in their computation of $a_\m^{\rm HVP}$.}   Assuming a linear relation between $m_\ell$ and $m_\p^2$, this form can be straightforwardly rewritten
in the form Eq.~(\ref{fitsc}).

\begin{figure}[t]
\vspace*{4ex}
\begin{center}
\includegraphics*[width=7.4cm]{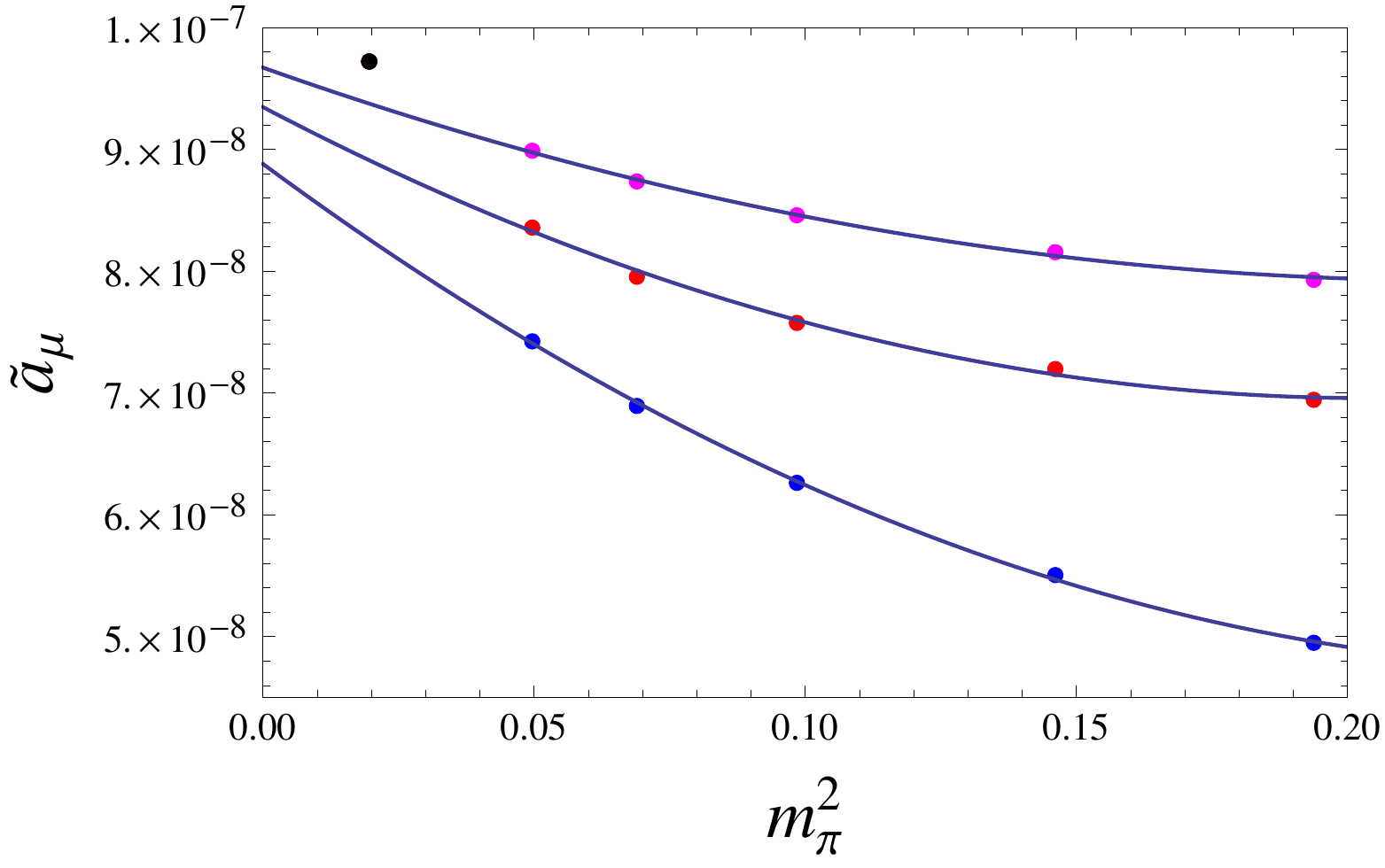}
\hspace{1ex}
\includegraphics*[width=7.4cm]{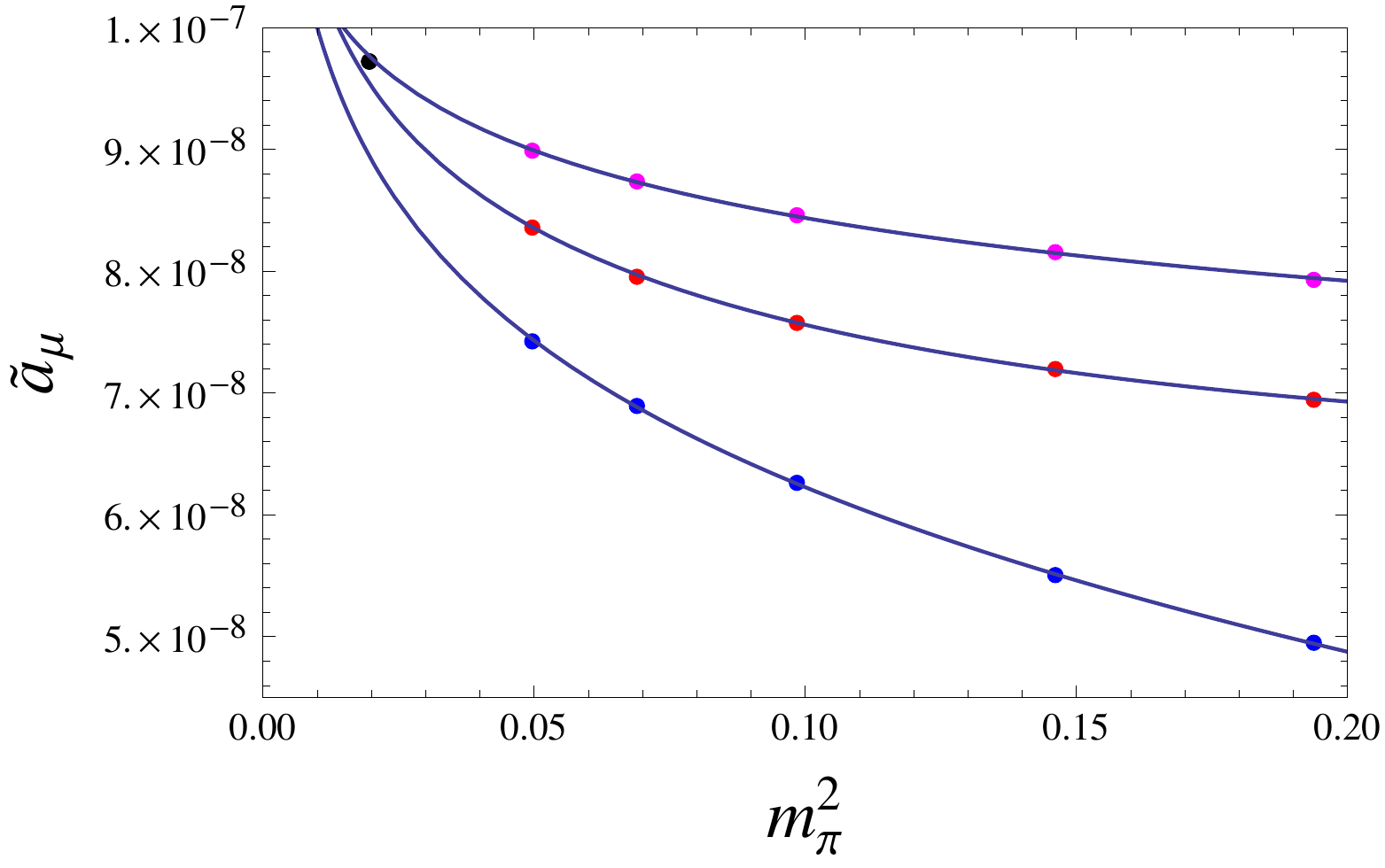}\\
\vspace*{4ex}
\includegraphics*[width=7.4cm]{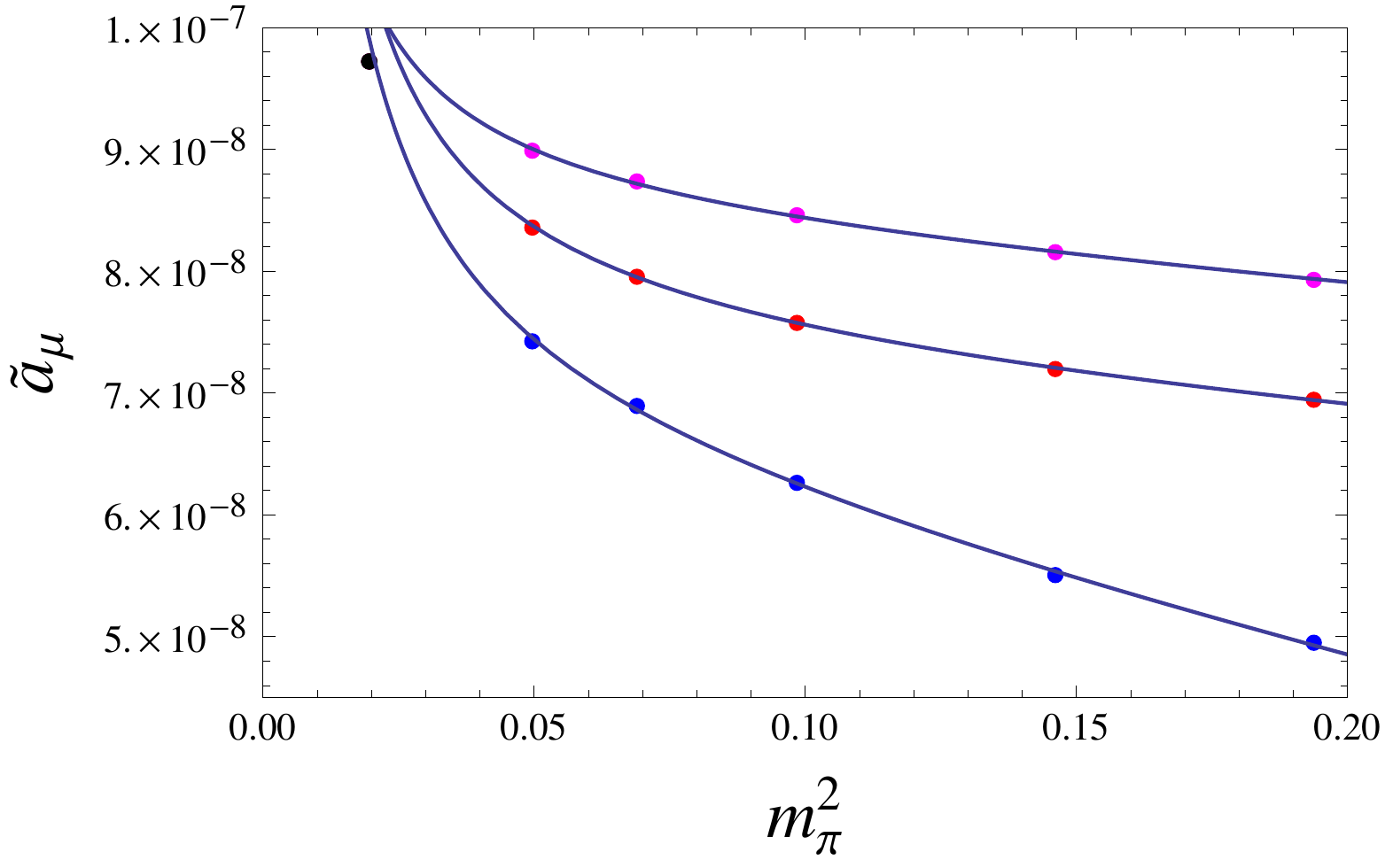}
\end{center}
\begin{quotation}
\floatcaption{fitfig}%
{\it The unmodified and ETMC- and HPQCD-improved versions of 
$\ta_\m$ as a function of $m_\p^2$. In each plot, the upper (magenta) data 
points are HPQCD-improved, the middle (red) data points ETMC-improved
and the lower (blue) data points unimproved.
The (black) point in the upper left corner of each plot is the 
``physical'' point, $\ta_\m=9.73\times 10^{-8}$ (\seef\ Eq.~(\ref{01})).
Fits are ``quadratic'' (upper left panel), ``log'' (upper right panel)
and ``inverse'' (lower panel). For further explanation, see text.}
\end{quotation}
\vspace*{-4ex}
\end{figure}

All three fits on all three data sets
are shown in Fig.~\ref{fitfig}.   Clearly, as expected,
the ETMC trick, and even more so the HPQCD trick, improve (\ie, reduce)
the pion mass dependence of the resulting modified $\ta_\m$: the values 
at larger pion masses are
closer to the correct ``physical'' value shown as the black point in the
upper left corner of all panels. All fits look good,\footnote{Of course,
in this study there are no statistical errors, and we can only judge 
this by eye.} and the log fits to the ETMC- or HPQCD-improved data
approach the correct value. However, the coefficient of the logarithm
in Eq.~(\ref{fitsb}) falls in the range 
$-1.5\times 10^{-8}$ to $-0.8\times 10^{-8}$,
more than an order of magnitude smaller than
the value $-4.5\times 10^{-7}$ predicted by Eq.~(\ref{amufinal}).   
In this respect, we note that the expansion~(\ref{amufinal})
has a chance of being reliable for $m_\p<m_\m$; the log fits, however, are
carried out for lattice pion masses which are 
larger than the physical pion mass, which,
in turn, is larger than $m_\m$.
We also note that the form~(\ref{fitsc})
is more singular than predicted by Eq.~(\ref{amufinal}).

We conclude that all three fits
are at best phenomenological, with none of the fit forms
in Eq.~(\ref{fits}) theoretically preferred. We also note that replacing the
linear term in $m_\p^2$ in Eq.~(\ref{fitsb}) by a term linear in $m_\p$, as
suggested by Eq.~(\ref{amufinal}), does not improve the mismatch between
the theoretical and fitted values of the coefficient of the logarithm.
All this suggests that the systematic error from the extrapolation to
the physical pion mass is hard to control, at least when the lowest
lattice pion mass is around $200$~MeV (or, when the statistical error
on a value closer to the physical pion mass is too large to 
sufficiently constrain the extrapolation).

\begin{table}[t]
\begin{center}
\vspace*{4ex}
\hspace{0cm}\begin{tabular}{|c|c|c|c|}
\hline
& unimproved data & ETMC-improved data  & HPQCD-improved data \\
\hline
quadratic & 8.26 & 8.91 & 9.38 \\
\hline
log & 8.96 & 9.55 & 9.77 \\
\hline
inverse & 9.93 & 10.46 & 10.33 \\
\hline
\end{tabular}
\vspace*{2ex}
\floatcaption{tab1}{\it Values for $\ta_\m\times 10^8$ for the three types of
fit (\seef\ Eq.~(\ref{fits})) and the three data sets. For reference, the correct
model value is $\ta_\m\times 10^8=9.73$ (\seef\ Eq.~(\ref{01})).}
\end{center}
\end{table}%

In Table~\ref{tab1} we show the values for $\ta_\m$ at the physical pion 
mass obtained from the three types of fit to the three data sets. The 
log-fit value to the HPQCD-improved data is particularly good, missing 
the correct value by only $0.4\%$. However, as we have seen, the log fit 
is not theoretically preferred, and without knowledge of the correct value, 
the only way to obtain an estimate for the systematic error associated 
with the extrapolation in the real world would be by comparing the results 
obtained using different fit forms. Discarding the inverse fit as 
too singular, one may take the (significantly smaller) variation 
between the quadratic and log fits as a measure of the systematic
error. This spread is equal to 8\%, 7\% and 4\%, respectively, for the
unimproved, ETMC-improved and HPQCD-improved data sets. Therefore,
even though the ETMC and HPQCD tricks do improve the estimated
accuracy, they are not sufficiently reliable to reach the desired level of 
sub-1\% accuracy.

We have also carried out the same fits omitting the highest
pion mass (of 440~MeV, \seef\ Eq.~(\ref{MILC})), and find this makes 
very little difference. The extrapolated values reported in 
Table~\ref{tab1} do not change by more than about $0.5$ to 1\%,
and there is essentially no change in the systematic uncertainty
estimated using the variation with the fit-form choice as we did
above.

\begin{boldmath}
\subsection{\label{EM} The electromagnetic case}
\end{boldmath}
We may repeat the analysis of Sec.~\ref{I=1} for the electromagnetic case,
\ie, using $\P_{\rm EM}(Q^2)$ instead of $\P^{{33}}(Q^2)$. The only
difference is that in this case we do not have a ``data'' value as in
Eqs.~(\ref{01}) and~(\ref{02}), and we have to rely on ChPT alone.

\begin{figure}[t]
\vspace*{4ex}
\begin{center}
\includegraphics*[width=7.4cm]{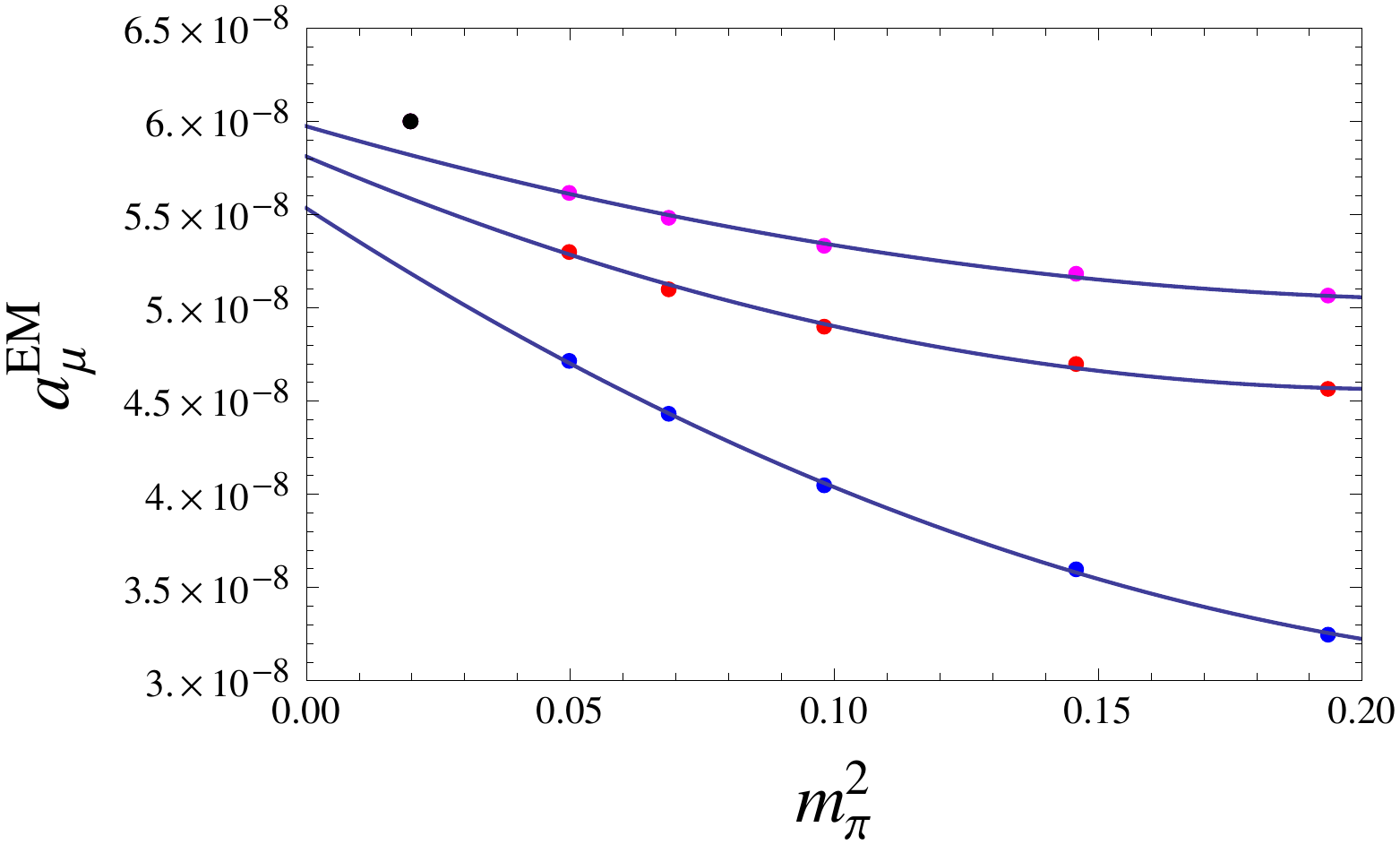}
\hspace{1ex}
\includegraphics*[width=7.4cm]{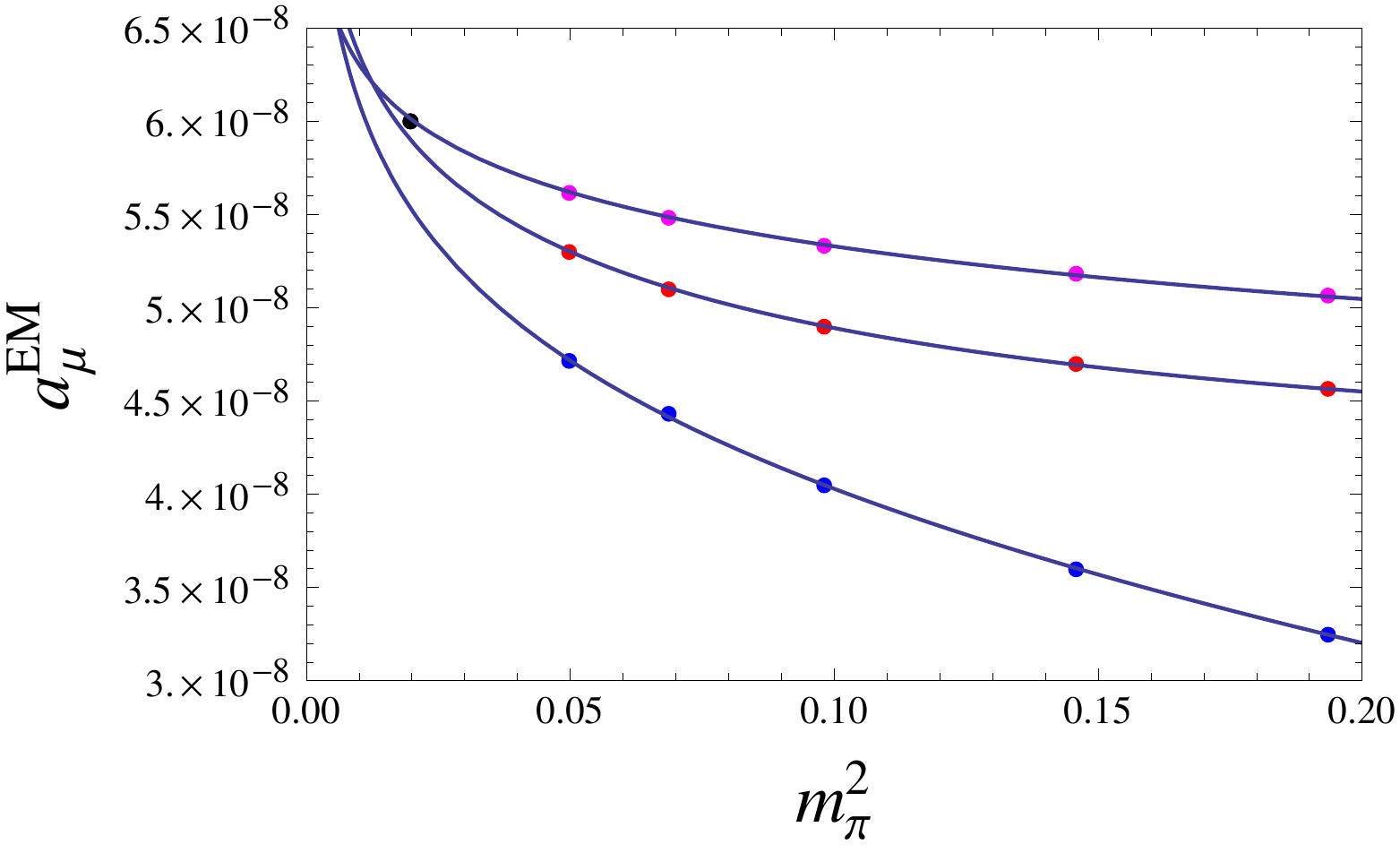}\\
\vspace*{4ex}
\includegraphics*[width=7.4cm]{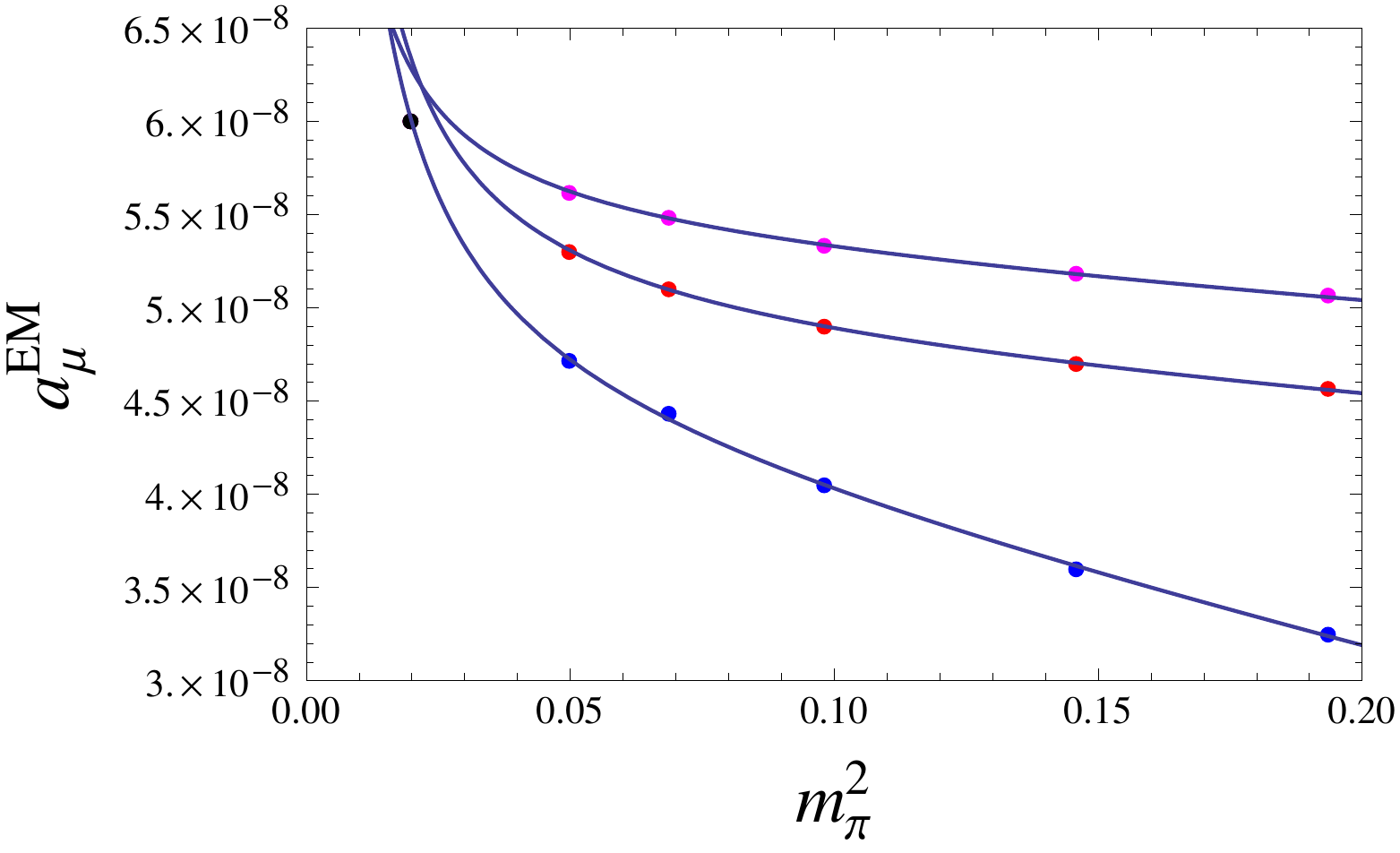}
\end{center}
\begin{quotation}
\floatcaption{fitfigEM}%
{\it The unmodified and ETMC- and HPQCD-improved versions of 
$a^{\rm EM}_\m$ as a function of $m_\p^2$. In each plot, the upper 
(magenta) data points are HPQCD-improved, the middle (red) data points 
ETMC-improved and the lower (blue) data points unimproved.
The (black) point in the upper left corner of each plot is the 
``physical'' point, $a^{\rm EM}_\m=6.00\times 10^{-8}$.
Fits are ``quadratic'' (upper left panel), ``log'' (upper right panel)
and ``inverse'' (lower panel). For further explanation, see text.}
\end{quotation}
\vspace*{-4ex}
\end{figure}

Defining the shorthand $a^{\rm EM}_\m=a^{\rm EM}_\m(Q^2_{max}=0.1$~GeV$^2)$,
and using the same notation for the ETMC- and HPQCD-modified versions
thereof, we show the quadratic, log and inverse fits for this case in 
Fig.~\ref{fitfigEM}. We see again that the use of the ETMC and HPQCD 
tricks significantly reduces the pion mass dependence of the resulting
modified data, with the HPQCD improvement being especially effective in 
this regard. The predicted value of the coefficient of the logarithm in the 
EM analogue of Eq.~(\ref{amufinal}) is now half the value shown in that equation 
(\seef\ Eq.~(\ref{EMVP})), equal to $-2.2\times 10^{-7}$, while the fitted 
coefficients for the log fits range between $-0.8\times 10^{-8}$ and 
$-0.4\times 10^{-8}$. The relative difference is of the same order as in 
the $I=1$ case, and the log fit should thus, as before, be considered 
purely phenomenological in nature.

\begin{table}[t]
\begin{center}
\vspace*{4ex}
\hspace{0cm}\begin{tabular}{|c|c|c|c|}
\hline
& unimproved data & ETMC-improved data  & HPQCD-improved data \\
\hline
quadratic & 5.19 & 5.59 & 5.82 \\
\hline
log & 5.55 & 5.91 & 6.02 \\
\hline
inverse & 6.04 & 6.37 & 6.30 \\
\hline
\end{tabular}
\vspace*{2ex}
\floatcaption{tab2}{\it Values for $a^{\rm EM}_\m\times 10^8$ for the three 
types of fit (\seef\ Eq.~(\ref{fits})) and the three data sets. For reference, 
the correct model value is $a^{\rm EM}_\m\times 10^8=6.00$.}
\end{center}
\end{table}%

In Table~\ref{tab2} we show the values for $a^{\rm EM}_\m$ at the physical 
pion mass obtained from the three types of fit to the three data sets.
The log-fit value to the HPQCD-improved data is particularly good, missing 
the correct value by only $0.3\%$. Discarding again the inverse fit as too 
singular, and taking the variation between the quadratic and log fits as a 
measure of the systematic error, the spread is 6\%, 5\% and 3\%, 
respectively, for the unimproved, ETMC-improved and HPQCD-improved data 
sets. Therefore, even though the ETMC and HPQCD tricks do improve the 
estimated accuracy, this improvement is not sufficient to reach 
the desired target of sub-1\% accuracy.  Again, removing the highest 
pion mass points from the fits makes no signficant difference in
these conclusions.

We conclude that, in the electromagnetic case, the situation is 
slightly better than in the $I=1$ case, no doubt because of the 
larger relative weight of contributions which are less sensitive to the 
pion mass (such as the two-kaon contribution).  While the ETMC and
HPQCD tricks do again improve the estimated accuracy, these
improvements remain insufficient to reliably reach the desired sub-1\% level.

\vskip 0.8cm
\section{\label{conclusion} Conclusion}
In this paper, we used a ChPT-inspired model to investigate the extrapolation
of the leading-order hadronic contribution to the muon anomalous magnetic 
moment, $a_\m^{\rm HLO}$, from lattice pion masses of order 200~to~400~MeV 
to the physical pion mass. We found that such pion masses are too large to 
allow for a reliable extrapolation, if the aim is an extrapolation error of 
less than 1\%. This is true even if various tricks to improve the 
extrapolation are employed, such as those proposed in Ref.~\cite{Fengetal} 
and Ref.~\cite{HPQCD}.   

In order to perform our study, we had to make certain
assumptions. First, we assumed that useful insight into the pion mass 
dependence could be obtained by focussing on the contribution to 
$a_\m^{\rm HLO}$ from $Q^2$ up to $Q^2_{max}=0.1$~GeV$^2$. This restriction 
is necessary if we want to take advantage of information on the
mass dependence from ChPT, since it is only in this range that ChPT 
provides a reasonable representation of the HVP. We believe this
is not a severe restriction, since that part of the integral yields 
over 80\% of $a_\m^{\rm HLO}$, and it is clear that it is the 
low-$Q^2$ part of the HVP which is most sensitive to the pion mass. 
Changing $Q^2_{max}$ to $0.2$~GeV$^2$ makes no qualitative difference 
to our conclusions.

Second, we assumed Eq.~(\ref{Ceff}) for the dependence of the
effective LECs $C^r_{93,{\rm eff}}$ and $C^r_{\rm eff}$ on the pion mass. 
While this is a phenomenological assumption, we note that this 
assumption is in accordance with the ideas underlying the ETMC and 
HPQCD tricks, so that those tricks should work particularly well if 
indeed this assumption would be correct in the real world. There are 
two reasons that the modified extrapolations nevertheless do not work 
well enough to achieve the desired sub-1\% accuracy.  One is the 
fact that in addition to the physics of the $\r$, the two-pion 
intermediate state contributing to the non-analytic terms in 
Eq.~(\ref{Pi1}), not just at one loop, but also beyond one loop, 
plays a significant role as well. This is especially so because of 
the structure of the weight function $w(Q^2)$ in Eq.~(\ref{amu}). 
The second reason is that, although  ChPT provides a simple functional form for the chiral extrapolation of $\tilde{a}_\m^{\rm HLO}$ for pion masses much smaller than the muon mass (\seef\ Eq.~(\ref{amufinal})), this is not useful in practice, so that one needs to rely on phenomenological fit forms, such as those of Eq.~(\ref{fits}).

In order to eliminate the systematic error from the chiral extrapolation, 
which we showed to be very difficult to estimate reliably, one  
needs to compute $a_\m^{\rm HLO}$ at, or close to, the
physical pion mass. This potentially increases systematic errors due to
finite-volume effects, but it appears these may be more easily brought
under theoretical control \cite{ABGPFV,Lehnerlat16,BRFV} than the
systematic uncertainties associated with a long extrapolation to the 
physical pion mass. Contrary to the experience with simpler quantities 
such as, \eg,  meson masses and decay constants, even an extrapolation 
from approximately 200~MeV pions turns out to be a long extrapolation.

It would be interesting to consider the case in which extrapolation 
from larger than physical pion masses is combined with direct computation 
at or very near the physical pion mass in order to reduce the total error 
on the final result. This case falls outside the scope of the study
presented here, because in this case the trade-off between extrapolation and
computation at the physical point is expected to depend on the statistical 
errors associated with the ensembles used for each pion mass. However, our 
results imply that also in this case a careful study should be made of the
extrapolation.  The methodology developed in this paper 
can be easily adapted to different pion masses and extended to take 
into account lattice statistics, and thus should prove very useful for 
such a study.

\vspace{3ex}
\noindent {\bf Acknowledgments}
\vspace{3ex}

We would like to thank Christopher Aubin, Tom Blum and Cheng Tu for 
discussions, and Doug Toussaint for providing us with unpublished
hadronic quantities obtained by the MILC collaboration.
This material is based 
upon work supported by the U.S. Department of Energy, Office of Science, 
Office of High Energy Physics, under Award Number DE-FG03-92ER40711
(MG). 
KM is supported by a grant from the Natural Sciences and Engineering Research
Council of Canada.  SP is supported by CICYTFEDER-FPA2014-55613-P, 2014-SGR-1450 and the CERCA Program/Generalitat de Catalunya.

\appendix
\section{\label{appendix} Chiral behavior of \boldmath$a_\m^{\rm HVP}$}
In this appendix, we derive the dependence of $a_\m^{I=1}$ on $m_\p$,
for $m_\p\to 0$ (in particular, $m_\p\ll m_\m$), using the lowest order pion-loop expression for the 
$I=1$ HVP, which we will denote by $\P^{33,{\rm NLO}}_{\rm sub}(Q^2)$.
Writing this as a dispersive integral,
\begin{eqnarray}
\label{disp}
\P^{33,{\rm NLO}}_{\rm sub}(Q^2)&=&
-Q^2\int_{4m_\p^2}^\infty\frac{dt}{t}\,\frac{\r^{33}(t)}{t+Q^2}\ ,\\
\r^{33}\left(\frac{4m_\p^2}{t}\right)&=&\frac{\a}{6\p}\left(1-\frac{4m_\p^2}{t}\right)^{3/2}\ ,\nonumber
\end{eqnarray}
and using Eq.~(\ref{amu}), the integral for $a_\m^{I=1}$ can be written as
\cite{EduardodeR}
\begin{subequations}
\label{amudisp}
\begin{eqnarray}
a_\m^{I=1}&=&\frac{\a}{\pi}\int_0^\infty\frac{d\o}{\o}\,f(\o)
\int_1^\infty\frac{d\t}{\t}\,\frac{\r^{33}(\t)}{1+\frac{\z\t}{\o}}\ ,\label{amudispa}\\
f(\o)&=&w(m_\m^2\o)=\sqrt{\frac{\o}{4+\o}}\left(\frac{\sqrt{4+\o}-\sqrt{\o}}{\sqrt{4+\o}+\sqrt{\o}}\right)\ ,
\label{amudispb}\\
\t&=&\frac{t}{4m_\p^2}\ ,\label{amudispc}\\
\z&=&\frac{4m_\p^2}{m_\m^2}\ .\label{amudispd}
\end{eqnarray}
\end{subequations}
Employing the Mellin--Barnes representation \cite{Greynat}
\begin{equation}
\label{MB}
\frac{1}{1+\frac{\z\t}{\o}}=\frac{1}{2\p i}\int_C ds\left(\frac{\z\t}{\o}\right)^{-s}
\G(s)\G(1-s)\ ,
\end{equation}
with $C$ a line parallel to the imaginary axis with $\Re(s)$ inside the
fundamental strip $0<\Re(s)<1$, we find an expression for $a_\m^{I=1}$
after performing the integrals over $\t$ and $\o$,
\begin{subequations}
\label{MBamu}
\begin{eqnarray}
a_\m^{I=1}&=&\frac{\a^2}{6\p}\,\frac{1}{2\p i}\int_Cds\,\z^{-s}M(s)\ ,\label{MBamua}\\
M(s)&=&3\cdot 4^{s-1}s(s-1)\,\frac{\G^2(s)\G(1-s)\G\left(\half+s\right)\G(-2-s)}{\G\left(\frac{5}{2}+s\right)}\ .
\label{MBamub}
\end{eqnarray}
\end{subequations}
The singular expansion consisting of the sum over all singular terms from a 
Laurent expansion around each of the singularities of $M(s)$ equals
\begin{equation}
\label{singexp}
M(s)\asymp\frac{1}{2s^2}+\frac{\log{2}-\frac{31}{12}}{s}+\frac{3\p^2}{4}\frac{1}{s+\half}
+O\left(\frac{1}{(s+1)^3}\right)\ .
\end{equation}
Using that 
\begin{equation}
\label{simple}
\frac{1}{2\p i}\oint ds\,\frac{\z^{-s}}{(s+a)^{k+1}}=\frac{(-1)^k}{k!}\,\z^a\log^k{\z}\ ,
\end{equation}
and closing the contour in Eq.~(\ref{MBamua}) to the left, we find 
\begin{equation}
\label{amuzeta}
a_\m^{I=1}=
\frac{\a^2}{12\p^2}\left(-\log{\z}
+2\log{2}-\frac{31}{6}+\frac{3\p^2}{2}\sqrt{\z}
+O\left(\z\log^2{\z}\right)\right)\ .
\end{equation}
Substituting the expression given in Eq.~(\ref{amudispd}) for $\z$ yields 
Eq.~(\ref{amufinal}).

\vspace{3ex}

\end{document}